\documentclass[a4paper,fleqn]{cas-sc}

\usepackage[authoryear]{natbib}
\usepackage{amsmath}
\usepackage{booktabs}
\usepackage{caption}

\def\tsc#1{\csdef{#1}{\textsc{\lowercase{#1}}\xspace}}
\tsc{WGM}
\tsc{QE}
\tsc{EP}
\tsc{PMS}
\tsc{BEC}
\tsc{DE}

\begin{document}
\let\WriteBookmarks\relax
\def\floatpagepagefraction{1}
\def\textpagefraction{.001}

\shorttitle{TEAM-SimHRA for Team-Level HRA}
\shortauthors{Xiao et~al.}

\title{TEAM-SimHRA: A Team-Based Simulation Framework for Human Reliability Analysis Using Multi-Agent Large Language Models}

\author[1,2]{Xingyu Xiao}[style=chinese]
\credit{Conceptualization, Methodology, Software, Formal analysis, Data curation, Visualization, Validation, Writing - original draft.}

\author[1,2]{Jiejuan Tong}[style=chinese]
\credit{Conceptualization, Formal analysis, Supervision, Writing - review and editing.}

\author[1]{Jingang Liang}[style=chinese]
\credit{Supervision, Writing - review and editing.}

\author[1]{Haitao Wang}[style=chinese]
\credit{Supervision, Writing - review and editing.}

\affiliation[1]{organization={Institute of Nuclear and New Energy Technology, Tsinghua University},
            city={Beijing},
            postcode={100084},
            country={China}}
\affiliation[2]{organization={National Key Laboratory of Human Factors Engineering},
            city={Beijing},
            postcode={100094},
            country={China}}

\begin{abstract}
Team-level failure in nuclear control rooms arises not from isolated operator error, but from emergent interaction dynamics — delayed diagnosis, suppressed dissent, and authority-driven error propagation — that conventional human reliability analysis methods are structurally unable to model. This study introduces TEAM-SimHRA, a multi-agent large language model simulation framework that reconceptualizes human reliability as an interaction-driven emergent property of control room teams rather than a static individual attribute. Unlike existing approaches that assign fixed error probabilities to predefined tasks, TEAM-SimHRA reproduces collective cognition, role-conditioned authority dynamics, and real-time communication suppression across temporally evolving accident progressions. Validated against the Three Mile Island (1979) and Chernobyl (1986) accidents — the two most extensively documented nuclear team failures — the framework achieves face-validity pass rates of 43.5\% and 52.6\% respectively, reproducing near-historical decision delay (134.8 vs. 138 min), perfect communication suppression stability, and full authority pressure cascade at historically accurate propagation depth. These results demonstrate that multi-agent simulation can extract quantitative team-level reliability indicators that are inaccessible to traditional methods, opening a viable path toward simulation-based dynamic probabilistic risk assessment for safety-critical sociotechnical systems.
\end{abstract}

\begin{keywords}
human reliability analysis \sep team cognition \sep nuclear control room \sep large language model \sep multi-agent simulation \sep authority pressure \sep communication suppression
\end{keywords}

\maketitle

\section{Introduction}

Nuclear accidents do not happen because a single operator makes a single mistake \cite{zhang2025analysis}. They happen because teams fail, through delayed diagnosis, suppressed dissent, authority-driven commitment to incorrect frames, and the progressive collapse of shared situational awareness under stress \cite{mccormick2025interpretive}. Yet the dominant paradigm of human reliability analysis (HRA) in probabilistic risk assessment remains built around individual operators, predefined task sequences, and static error probabilities \cite{bovo2020detecting}. Methods such as THERP, SPAR-H, and IDHEAS have made indispensable contributions to structured safety analysis, but their architectural assumptions preclude representation of the interaction dynamics through which team-level failure actually emerges \cite{kozlowski2018unpacking}.

The evidence from major accidents is unambiguous. At Three Mile Island (1979) \cite{sills2019accident}, the crew collectively locked onto an incorrect system interpretation for nearly 138 minutes, not because any individual lacked competence, but because incomplete communication and shared cognitive inertia prevented the team from updating its frame despite contradictory cues. At Chernobyl (1986) \cite{healey2015teams}, operators continued executing a fatally flawed test procedure not from ignorance alone, but because hierarchical authority suppressed dissent at every level of the team structure. In both cases, the accident trajectory was shaped by emergent interaction properties \cite{dekker2017safety}, communication patterns, authority gradients, and evolving collective mental models, that no individual-task decomposition can adequately represent. This is not a peripheral gap \cite{gupta2022transactive}. It is a structural limitation at the core of how HRA currently models human performance in safety-critical operations \cite{liu2021human}.

Recent advances in large language models (LLMs) and multi-agent simulation offer a fundamentally new approach to this problem \cite{gao2024large}. LLM-based agents can sustain role-conditioned behavior, generate context-sensitive dialogue, and reproduce sequential social interaction over extended temporal horizons. These capabilities align directly with what team-level HRA requires: a computational substrate that can represent operators not as isolated error sources but as communicating, authority-sensitive, cognitively interdependent agents operating within a shared and evolving plant context \cite{eloy2023capturing}. Nevertheless, existing LLM applications in safety-critical domains remain largely qualitative, and no prior framework has demonstrated that multi-agent simulation can produce quantitative HRA indicators that hold up against rigorous historical validation \cite{kiasari2026agentic}.

This paper addresses that gap by proposing TEAM-SimHRA, a team-based simulation framework that reconceptualizes human reliability as an emergent property of role-conditioned multi-agent interaction. The framework makes three principal contributions \cite{dang2025multi}. First, it operationalizes team cognition and authority dynamics as simulatable agent behaviors rather than fixed modifiers \cite{fan2004modeling}, enabling dynamic rather than static reliability representation. Second, it introduces five quantitative team-level HRA metrics, decision delay time, incorrect procedure rate, communication suppression rate, authority pressure cascade, and frame lock index, that capture failure mechanisms invisible to conventional task-based analysis \cite{wen2025survey}. Third, it provides the first systematic face-validity evaluation of multi-agent LLM simulation against two of the most thoroughly documented nuclear accident benchmarks, demonstrating that simulation-derived metrics can reproduce historically meaningful team failure patterns with quantifiable fidelity \cite{fridley2025autonomous}. Together, these contributions open a path toward simulation-based dynamic HRA that treats team interaction, not individual error, as the primary unit of reliability analysis \cite{groth2019hybrid}.

\section{Related Work}

\subsection{Human Reliability Analysis}
Human reliability analysis has evolved through several distinct methodological generations, each extending the representational capacity of its predecessors while preserving a common architectural assumption: that human performance can be adequately modeled at the level of the individual operator executing a predefined task. First-generation approaches such as THERP \cite{swain1964therp} decompose operator behavior into binary success-failure events and assign nominal error probabilities modulated by tabulated performance shaping factors. Second-generation methods, including ATHEANA \cite{cooper1996technique}, CREAM \cite{he2008simplified}, and SPAR-H \cite{whaley2012spar}, extended this foundation by incorporating error-forcing conditions and richer contextual representations. More recent frameworks such as IDHEAS \cite{chang2021idheas} have grounded HRA in cognitive task models derived from information processing theory, further improving the realism of individual operator representation. Each generation has meaningfully advanced the field, yet none has departed from the fundamental unit of analysis: a single operator, a single task, a single error probability.

The cognitive modeling embedded in these methods reflects a similarly constrained view of human performance under stress. Performance shaping factors, workload, time pressure, experience, procedure quality, enter the analysis as fixed multiplicative modifiers applied 
to nominal error rates, rather than as dynamic variables that interact and evolve across the course of an event \cite{mulder2019modeling}. 

This static formulation cannot represent the feedback dynamics through which cognitive states actually develop in abnormal situations: situational awareness degrades as cues accumulate and are selectively attended to, mental models update or fail to update in response to contradictory evidence, and stress reshapes both individual cognition and the social processes through which team members share and interpret information \cite{janhonen2011role}. The gap between 
the cognitive reality of control-room performance and the cognitive assumptions embedded in standard HRA methods is therefore not a matter of parameterization but of model architecture.

Most consequentially for the present study, established HRA frameworks provide no native representation of team-level interaction as a reliability-relevant process. Communication patterns, authority gradients, and the suppression of dissenting interpretation are 
treated, if at all, as background contextual factors rather than as dynamic mechanisms that actively shape accident trajectories \cite{factor2007social}. Yet the empirical record of major nuclear accidents consistently implicates precisely these mechanisms as primary 
drivers of failure propagation, a point developed in the following subsection. The absence of team interaction from the core modeling apparatus of HRA therefore represents a structural limitation that cannot be addressed through incremental refinement of existing methods alone.

\subsection{Team dynamics and communication failure in nuclear accidents}

A substantial body of accident investigation and organizational safety research has established that team-level interaction, rather than isolated individual error, is the primary driver of failure propagation in high-stakes operational environments. Reason's organizational accident model \cite{reason2016organizational} identifies latent conditions embedded in team structure and communication norms as the deeper causes of catastrophic failure, while Hollnagel's cognitive systems engineering 
perspective \cite{militello2010role} further demonstrates that performance breakdown in complex sociotechnical systems cannot be reconstructed from the sum of individual contributions. In nuclear control rooms specifically, empirical analyses of abnormal event responses consistently show that crew outcomes are determined less by the technical competence of individual operators than by the quality of shared situational awareness, the openness of communication channels, and the distribution of decision authority within the team 
\cite{seppanen2013developing, marusich2016effects}.

The two accident scenarios examined in this study illustrate these mechanisms with particular clarity. At Three Mile Island, the operators possessed sufficient individual knowledge to recognize the loss-of-coolant signature had the available cues been correctly integrated and communicated across the team. Instead, an incorrect 
diagnostic frame, that the primary system was fully pressurized, became collectively entrenched within the first minutes of the event and resisted correction for nearly 138 minutes despite accumulating contradictory evidence \cite{vanini2024clock}. The failure 
was not one of individual cognition but of collective frame lock: a team-level phenomenon in which shared commitment to an initial interpretation suppresses the uptake of disconfirming information. 
At Chernobyl, the mechanism was different but equally team-level in character. Deputy Chief Engineer Dyatlov's authority over the shift created a hierarchical pressure gradient that propagated downward through Akimov to Toptunov, systematically suppressing the expression 
of operational concern at each level despite reactor behavior that contradicted the test assumptions \cite{pastore2015uncertainty}. 

In both cases, the accident trajectory was shaped not by what any individual operator knew or failed to know, but by how the team communicated, deferred, and collectively committed to a course of action.

These observations point to a set of team-level failure mechanisms that recur across major nuclear accidents and that have been systematically characterized in the human factors and organizational safety literature: delayed challenge of authority, incomplete escalation of anomalous cues, collective commitment to an incorrect diagnostic frame, and the progressive silencing of dissent under 
hierarchical pressure \cite{seitz2025quiet}. What is significant for HRA is not merely that these mechanisms exist, but that they are quantifiable in principle, communication suppression has a rate, authority pressure has a propagation depth, diagnostic delay has a duration, and that their quantification would constitute genuinely new information for risk assessment. The challenge is that 
no existing HRA framework provides a computational substrate capable of generating these quantities from a model of team interaction. Addressing that challenge requires a fundamentally different modeling approach, as the following subsection discusses.

\subsection{Large language model agents for sociotechnical simulation}

The emergence of large language models capable of sustained, role-conditioned interaction has created qualitatively new possibilities for simulating human behavior in complex social contexts. Early applications demonstrated that LLM-based agents can reproduce negotiation dynamics \cite{mouri2025simulating}, 
coordinate collaborative planning under uncertainty 
\cite{chakraborty2012recognizing}, and generate emergent social phenomena in multi-agent environments without explicit behavioral programming \cite{cardoso2021review}. More recent work has extended these capabilities to organizational role simulation, showing that 
LLM agents assigned distinct authority levels, expertise boundaries, and behavioral dispositions can sustain role-consistent interaction 
across extended dialogue sequences while responding coherently to externally injected state changes \cite{traum1999model}. 
These properties, role stability, context sensitivity, and 
sequential interaction coherence, map directly onto the requirements of control-room team simulation, where operators must maintain role-defined behavioral tendencies while processing a continuously evolving plant state.

Applications of LLM-based simulation in safety-critical and engineering domains have begun to accumulate. In nuclear human factors specifically, recent work has applied neuro-symbolic LLM architectures to procedural execution in digital control rooms \cite{kim2025bridging}, integrated knowledge graphs with LLM reasoning for base rate HRA \cite{shen2026reason}, and coupled cognitive architecture models with uncertainty quantification 
platforms for dynamic reliability assessment 
\cite{kabir2018uncertainty}. Beyond nuclear applications, LLM agents have been used to model decision-making in emergency response \cite{xiao2025novel}, organizational communication breakdown \cite{damian2007awareness}, and crew coordination in aviation contexts \cite{grote2010adaptive}. Across these applications, a consistent finding emerges: LLM agents can generate behaviorally plausible interaction trajectories that reflect domain-specific role constraints and social dynamics in ways that rule-based simulation cannot easily replicate.

Nevertheless, a critical methodological gap remains. Existing applications in safety-critical domains share two limitations that constrain their contribution to formal HRA. First, simulation outputs are evaluated qualitatively or against synthetic scenarios, without systematic comparison to documented historical accident data that 
could establish face validity at the level required for probabilistic risk assessment \cite{bui2024probabilistic}. Second, and more fundamentally, no existing framework translates multi-agent interaction trajectories into quantitative team-level HRA indicators, the kind of structured, numerically expressed reliability metrics that can be integrated into dynamic probabilistic risk assessment workflows. Generating plausible narratives is not equivalent to 
producing actionable reliability quantities. What the field requires is a framework that closes both gaps simultaneously: one that extracts quantitative team-level metrics from simulated interaction and validates those metrics against the historical accident record. TEAM-SimHRA is designed to meet precisely that requirement.

\section{Methodology}
\label{sec:method}

\subsection{Framework overview}

TEAM-SimHRA models the nuclear control room as a dynamic multi-agent 
system in which human reliability emerges not from individual operator 
characteristics alone, but from the interaction structure of the team 
under continuously evolving plant conditions. This design reflects a 
core theoretical commitment: that team-level failure mechanisms such 
as communication suppression, authority pressure cascade, and 
collective frame lock cannot be reproduced by aggregating individual 
error models, but only by simulating the interaction process through 
which they arise. The framework is accordingly organized around five 
functionally distinct components whose coordinated operation produces 
a temporally structured simulation of control-room team behavior.

The first component, \textit{role-conditioned operator agents}, 
instantiates each crew member as an independent LLM agent whose 
behavior is governed by a role prompt encoding domain expertise, 
operational responsibility, authority position within the team 
hierarchy, and characteristic behavioral tendencies under abnormal 
conditions. This role conditioning is the primary mechanism through 
which the framework differentiates operator behavior without requiring 
separate model architectures for each agent. The second component, 
the \textit{shared dialogue state}, functions as the public 
communication channel of the control room: all agent utterances are 
appended to a common buffer that every agent reads before generating 
its next response, ensuring that team interaction is grounded in a 
genuinely shared and sequentially coherent information environment. 
The third component, \textit{world event injection}, updates the 
external plant state at each simulation round according to a 
predefined historical timeline, providing the evolving operational 
context to which agents must respond. Together, these three components 
constitute the observable simulation environment.

The remaining two components operate at a supervisory level. The 
\textit{moderator agent} monitors agent outputs at the end of each 
round and, when behavioral drift is detected, such as premature escalation, historically implausible recovery, or role-inconsistent 
action, injects hidden corrective guidance into the role context of 
the relevant agent for the subsequent round. Critically, this 
guidance is not visible in the public dialogue and therefore does not 
alter the observable interaction record; it functions as a lightweight 
behavioral guardrail that preserves historical plausibility without 
contaminating the simulated communication channel. The \textit{report 
agent} then operates post-hoc on the complete interaction log, 
parsing dialogue trajectories to extract five structured team-level 
HRA metrics that form the quantitative output of the framework. The 
overall simulation pipeline is formalized as

\begin{equation}
    S_0 \xrightarrow{\text{event injection}} S_t 
    \xrightarrow{\text{agent interaction}} D_t 
    \xrightarrow{\text{moderation}} D_{t+1} 
    \xrightarrow{\text{report agent}} M,
\end{equation}

\noindent where $S_t$ denotes the evolving scenario state at round $t$, $D_t$ denotes the accumulated dialogue history, and $M$ denotes the extracted metric set. The architecture is illustrated in Figure~\ref{fig:framework}.

\begin{center}
\includegraphics[width=0.8 \textwidth]{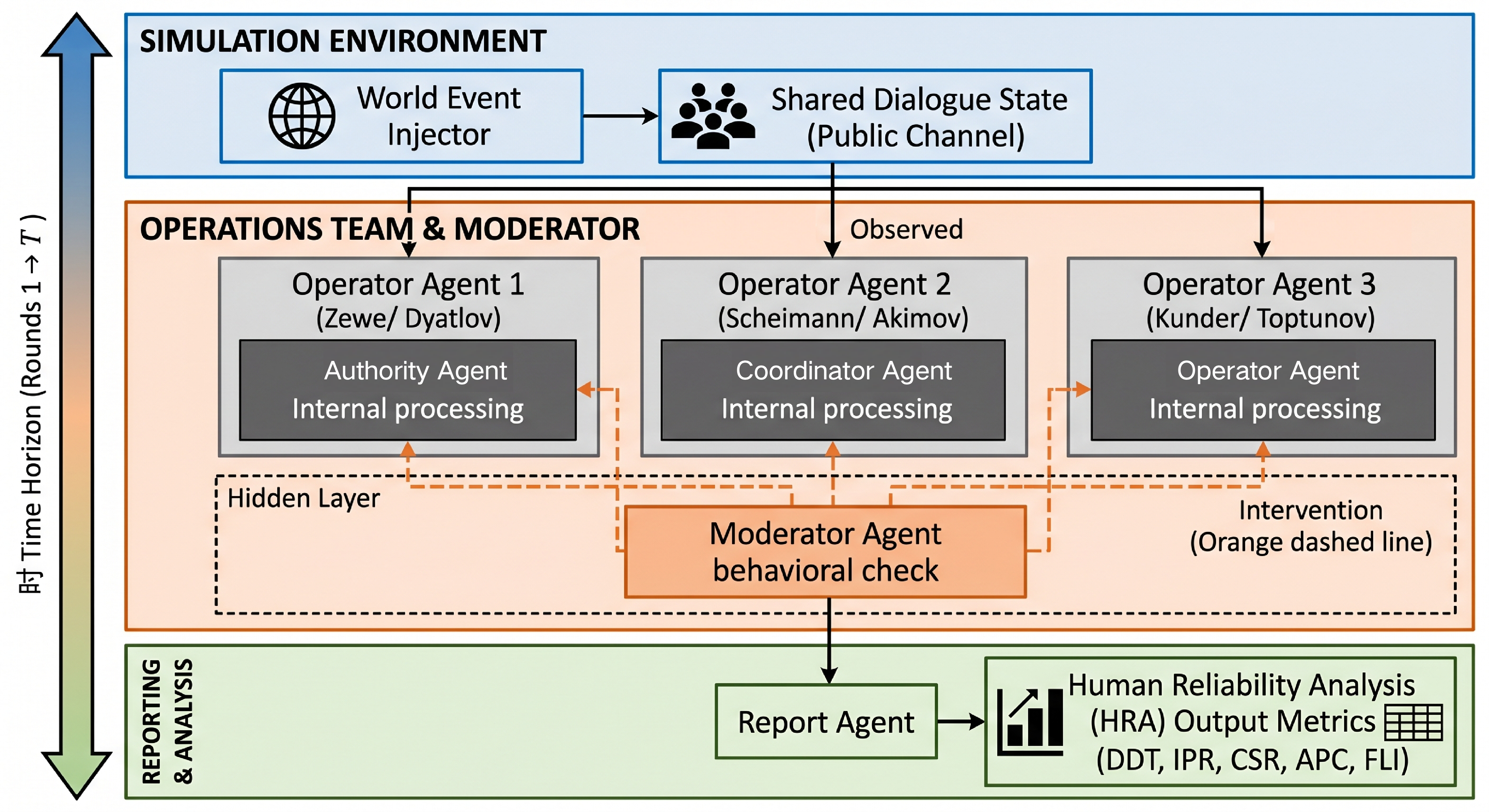}
\captionof{figure}{Multi-Agent Team Interaction Simulation Framework.}
\label{fig:framework}
\end{center}

\subsection{Operator representation}

Each operator in TEAM-SimHRA is instantiated as an independent LLM agent whose behavior is governed by a structured role prompt rather than by separate model architectures or fine-tuned weights. The role prompt encodes four functionally distinct dimensions: 
\textit{knowledge boundary}, which defines the operator's domain expertise and the informational scope within which their judgments are considered authoritative; \textit{operational responsibility}, which specifies the procedural tasks and system domains for which the agent is accountable; \textit{authority position}, which establishes the agent's rank within the team hierarchy and 
determines both their capacity to issue directives and their susceptibility to hierarchical pressure from above; and \textit{stress-conditioned behavioral tendencies}, which encode characteristic response patterns under abnormal conditions, including propensity for assertive challenge, deference to authority, or commitment to an established diagnostic frame. 
This four-dimensional role formulation is the primary mechanism through which behaviorally differentiated agents are produced from a single shared model backbone.

The role assignments for each accident scenario are grounded in historical documentation of the actual crew compositions and interpersonal dynamics present during the events. Across 
both scenarios, three agents are instantiated corresponding 
to a three-tier authority hierarchy: the \textit{Authority Agent}, 
who holds directive primacy and is primarily responsible for 
overall diagnostic framing and decision commitment; the 
\textit{Coordinator Agent}, who occupies the intermediate 
authority position and functions as the primary channel for 
instruction relay and information filtering between the 
authority and operational levels; and the \textit{Operator Agent}, 
who holds the lowest authority position and bears direct 
system interface responsibility, rendering this agent most 
susceptible to hierarchical compliance pressure from above. 
The specific historical personnel instantiated in each role 
across the two scenarios are summarized in 
Table~\ref{tab:agents}.

\begin{table}[htbp]
\centering
\caption{Role assignments for simulated operator agents 
across the two accident scenarios.}
\label{tab:agents}
\begin{tabular}{lllll}
\toprule
\textbf{Agent Role} & \textbf{Authority} & 
\textbf{TMI-1979} & \textbf{Chernobyl-1986} & 
\textbf{Key Behavioral Tendency} \\
\midrule
Authority Agent   & High   
& Zewe (Shift Supervisor)          
& Dyatlov (Deputy Chief Engineer)  
& Frame-setting, dissent suppression \\
Coordinator Agent & Medium 
& Scheimann (Control Room Operator) 
& Akimov (Shift Foreman)           
& Procedure relay, pressure transmission \\
Operator Agent    & Low    
& Kunder (Senior Reactor Operator)  
& Toptunov (Reactor Operator)      
& System interface, hierarchical compliance \\
\bottomrule
\end{tabular}
\end{table}

This role-conditioned formulation serves two purposes 
simultaneously. At the individual level, it allows the same 
model backbone to express qualitatively distinct interaction 
styles, decision tendencies, and authority sensitivities across 
agents without requiring separate training or architectural 
differentiation. At the team level, it instantiates a structured 
authority hierarchy whose gradient directly determines the 
direction and depth of authority pressure cascade, the 
propagation of directive influence from Authority Agent through 
Coordinator Agent to Operator Agent that constitutes one of 
the five core HRA metrics extracted by the framework. The 
fidelity of this representation to historical crew structure 
is therefore not merely a matter of narrative authenticity 
but a prerequisite for the validity of the authority-related 
metrics that the framework is designed to produce.

\subsection{Interaction environment and temporal modeling}

The interaction environment of TEAM-SimHRA is organized around 
a shared dialogue buffer that functions as the public communication 
channel of the simulated control room. All agent utterances are 
appended to this common buffer in the order they are generated, 
and every agent reads the complete accumulated dialogue history 
before producing its next response. This architecture ensures 
that team interaction is grounded in a genuinely shared and 
sequentially coherent information environment: no agent has 
access to information that has not passed through the public 
channel, and no agent can act without awareness of what has 
previously been said. The shared buffer therefore serves not 
only as a communication medium but as the operationalization 
of collective situational awareness within the simulation,
its contents at any given round represent the totality of 
information that the team has collectively made explicit.

The simulation proceeds in discrete rounds, each representing a fixed interval of scenario time and containing one complete sequential interaction cycle in which all agents respond in turn. At the start of each round, the world event injector appends scenario-specific plant developments to the shared buffer, updating the observable operational context to which agents must respond. These injected events follow a predefined 
historical timeline derived from accident documentation, 
ensuring that the external stimulus sequence encountered by the simulated team reflects the actual progression of plant 
state during the historical event. Agents then respond 
sequentially, Authority Agent first, followed by Coordinator Agent and Operator Agent, using three inputs simultaneously: their current role prompt, the full accumulated dialogue 
history, and any hidden moderator guidance issued for that 
round. The temporal granularity of each round is calibrated 
separately for each scenario to align the simulation pace 
with the historical accident timeline, as summarized in 
Table~\ref{tab:temporal}.

\begin{table}[htbp]
\centering
\caption{Temporal configuration of TEAM-SimHRA across 
the two accident scenarios.}
\label{tab:temporal}
\begin{tabular}{lccc}
\toprule
\textbf{Scenario} & \textbf{Total Rounds} & 
\textbf{Time per Round} & \textbf{Total Simulated Duration} \\
\midrule
TMI-1979       & 15 & $\approx$10 min & $\approx$150 min \\
Chernobyl-1986 & 12 & $\approx$2 min  & $\approx$24 min  \\
\bottomrule
\end{tabular}
\end{table}

The difference in temporal granularity between the two scenarios 
reflects the fundamentally different dynamics of each accident. The Three Mile Island event unfolded over a prolonged diagnostic period in which the crew's collective frame remained locked for nearly 138 minutes despite accumulating contradictory evidence, a process that requires sufficient round resolution to capture 
the gradual reinforcement of an incorrect interpretation. The Chernobyl test sequence, by contrast, involved a rapid cascade of safety-critical decisions compressed into approximately 24 minutes, demanding finer temporal granularity to represent the speed at which authority pressure and procedural commitment propagated through the team hierarchy. Calibrating round duration to scenario-specific historical pace is therefore not merely a simulation convenience but a validity requirement: 
it ensures that the temporal structure of simulated team 
interaction is commensurable with the historical timeline 
against which face validity is subsequently assessed.

\subsection{Moderator-guided behavioral control}

A fundamental challenge in long-horizon multi-agent role 
simulation is behavioral drift: the tendency of LLM agents 
to deviate progressively from their assigned roles, historical framing, or scenario-appropriate response patterns as the dialogue accumulates. In the context of accident simulation, drift manifests in several characteristic forms. Agents may escalate prematurely, initiating recovery actions before the historical timeline would support them and thereby short-circuiting the very failure mechanisms the simulation is designed to reproduce. They may abandon historically grounded diagnostic frames in favor of more rationally optimal interpretations that real operators demonstrably did not reach under the conditions of the event. They may also generate role-inconsistent authority behavior, an Operator Agent challenging the Authority Agent assertively in a scenario where hierarchical compliance was historically total, for instance, that undermines the validity of authority-sensitive metrics such as the authority pressure cascade and communication suppression rate. Left unaddressed, behavioral drift does not merely reduce narrative plausibility; it systematically biases the HRA metrics that the framework is designed to extract.

Table~\ref{Moderator} reports moderator intervention 
statistics across both scenarios, disaggregated by 
drift type and primary round of occurrence. Several 
patterns are diagnostically informative. In the Three 
Mile Island scenario, rational override is the most 
frequently triggered intervention category, accounting 
for 22.5\% of all moderator activations and concentrated 
in rounds 9--12 — the phase of the simulation 
corresponding to the period in which the actual TMI 
crew was receiving the strongest contradictory evidence 
yet remained committed to the incorrect pressurizer 
interpretation. This pattern indicates that without 
moderator intervention, agents would have updated their 
diagnostic frames more rapidly than the historical 
record supports, confirming that the pressure toward 
rational override is an intrinsic tendency of current 
LLM role-playing behavior rather than an artifact of 
specific prompt formulations. Premature escalation 
accounts for 18.1\% of TMI interventions and occurs 
earlier in the timeline (rounds 4--8), reflecting 
agents' tendency to initiate recovery actions before 
the historical diagnostic delay has elapsed. Authority 
inversion is the least frequent intervention category 
in TMI (8.9\%), consistent with the less rigidly 
hierarchical team structure of that scenario relative 
to Chernobyl.

In the Chernobyl scenario, the intervention profile 
is markedly different. Premature escalation is nearly 
absent (1.8\%), reflecting the strong authority 
conditioning embedded in the Dyatlov role prompt that 
prevents lower-authority agents from initiating 
unsanctioned recovery actions. Rational override 
remains the dominant category (13.9\%), concentrated 
in rounds 8--11 as reactor behavior increasingly 
contradicted test assumptions, while authority 
inversion accounts for 7.9\% of interventions in 
rounds 7--9. The substantially lower overall 
intervention rate in Chernobyl relative to Three 
Mile Island is consistent with the sharper authority 
gradient and more compressed temporal dynamics of 
that scenario: the hierarchical role conditioning 
alone is more effective at sustaining historically 
plausible behavior when authority structure is 
unambiguous, reducing the moderator's corrective 
burden. Taken together, these statistics confirm 
that the moderator is not operating as a dominant 
behavioral script but as a targeted correction 
mechanism whose activation pattern itself reflects 
meaningful differences in the team dynamics of the 
two accidents.

\begin{table}[htbp]
\centering
\caption{Moderator intervention statistics across both scenarios.}
\begin{tabular}{lccc}
\toprule
\textbf{Scenario} & \textbf{Drift Type} & 
\textbf{Intervention Rate} & 
\textbf{Primary Round} \\
\midrule
TMI-1979      & Premature escalation & 18.1\% & Round 4--8  \\
TMI-1979      & Rational override    & 22.5\% & Round 9--12 \\
TMI-1979      & Authority inversion  &  8.9\% & Round 9--13 \\
Chernobyl-1986 & Premature escalation &  1.8\% & Round 5--9  \\
Chernobyl-1986 & Rational override    & 13.9\% & Round 8--11 \\
Chernobyl-1986 & Authority inversion  &  7.9\% & Round 7--9  \\
\bottomrule
\end{tabular}\label{Moderator}
\end{table}

To address this challenge, TEAM-SimHRA introduces a moderator agent that operates as a dedicated supervisory layer outside the observable simulation environment. At the end of each round, the moderator receives the complete dialogue output of that round and evaluates each agent's utterances against three criteria: role consistency, which assesses whether the agent's behavior remains within the authority, expertise, and dispositional boundaries encoded in its role prompt; historical plausibility, which assesses whether the agent's actions and 
interpretations are consistent with what is documented of the 
corresponding historical figure at the equivalent stage of 
the accident; and scenario stage appropriateness, which 
assesses whether the agent's level of concern, diagnostic 
certainty, or recovery orientation is commensurate with the 
plant state and elapsed scenario time. When a violation of 
any criterion is detected, the moderator generates a hidden 
corrective note that is injected into the private context 
of the relevant agent for the subsequent round, redirecting 
behavior without altering the public dialogue record.

The design of the moderator agent reflects a deliberate 
methodological choice about the appropriate boundary between 
behavioral control and simulation authenticity. The moderator's 
corrective guidance is strictly non-public: it is never 
appended to the shared dialogue buffer and is therefore 
invisible to other agents and absent from the interaction 
log that the report agent subsequently parses. This separation 
ensures that the observable simulation record, the sequence 
of utterances that constitutes the primary data for metric 
extraction, reflects genuine agent interaction rather than 
externally scripted behavior. In effect, the moderator 
functions analogously to the implicit social and institutional 
constraints that shape real operator behavior in control rooms: present and consequential, but not directly observable in the communication record itself. The moderator thus serves not only as an engineering mechanism for maintaining scenario fidelity, but as a structural component whose design embodies the framework's commitment to producing historically faithful digital twins of control-room team dynamics.

\subsection{Metric extraction for team-level HRA}

Upon completion of each simulation run, a dedicated report agent parses the full interaction log and extracts five team-level HRA indicators that together constitute the quantitative output of the framework. The design of these indicators reflects a deliberate departure from conventional HRA metrics: rather than quantifying the probability of error on a predefined individual 
task, each indicator targets an emergent team-level phenomenon 
that arises from interaction and that has been implicated in 
historical accident progression. The report agent evaluates 
each indicator through structured prompting that directs the 
LLM to assess specific dialogue features — temporal patterns, 
utterance content, authority relationships, and interpretive 
commitments — and returns results in a structured JSON format 
that enables automated aggregation across repeated runs. The 
five indicators are defined as follows.

\textit{Decision delay time} (DDT) measures the elapsed scenario 
time between the onset of the abnormal event and the first 
utterance in which a team member correctly identifies the 
recovery-oriented action or diagnosis required by the situation. 
DDT operationalizes the collective diagnostic inertia of the 
team: a prolonged DDT indicates that the team as a whole failed 
to converge on a correct interpretation despite available 
evidence, regardless of whether any individual agent possessed 
the requisite knowledge. In scenarios where no correct recovery 
action is identified within the simulation horizon, DDT is 
recorded as NO\_RECOVERY, reflecting persistent non-recovery 
as a team-level outcome.

\textit{Incorrect procedure rate} (IPR) measures the proportion 
of procedure-related decisions or utterances within the 
interaction log that are assessed as incorrect, misleading, 
or inconsistent with the normatively required response at 
that stage of the scenario. IPR captures the degree to which 
the team's collective procedural execution deviates from 
correct practice, a deviation that may arise from individual 
misunderstanding, from authority-driven suppression of 
corrective input, or from the propagation of an incorrect 
diagnostic frame through the team hierarchy.

\textit{Communication suppression rate} (CSR) measures the 
degree to which critical challenges, safety-relevant concerns, 
or dissenting interpretations are suppressed within the team 
dialogue. An utterance is classified as suppressed when a 
team member either fails to voice a concern that their role 
and knowledge would warrant, or when a voiced concern is 
dismissed, ignored, or overridden without substantive 
engagement. CSR therefore captures the social filtering 
process through which potentially corrective information 
is prevented from influencing collective decision-making, a mechanism central to both the Three Mile Island and 
Chernobyl accident narratives.

\textit{Authority pressure cascade} (APC) assesses whether 
directive influence from the Authority Agent propagates 
downward through the Coordinator Agent to the Operator Agent in a manner that constrains the lower-level agents' ability to exercise independent judgment. APC is evaluated along two dimensions: cascade presence, recorded as a binary indicator of whether the propagation pattern is observed; and cascade depth, recorded as the number of hierarchical levels through which directive pressure demonstrably transmitted. A cascade depth of two indicates full propagation from Authority Agent through Coordinator Agent to Operator Agent, corresponding to the historically 
documented authority structure in both benchmark scenarios.

\textit{Frame lock index} (FLI) measures the intensity of 
the team's persistent commitment to an incorrect interpretive frame despite the availability of disconfirming evidence. 
FLI is scored on an ordinal scale by the report agent, 
which assesses the frequency and consistency with which 
agents return to, reinforce, or act upon an incorrect 
diagnosis across successive rounds even as contradictory 
plant cues accumulate. A higher FLI indicates stronger 
collective resistance to frame revision, the team-level 
analog of confirmation bias that was particularly pronounced in the Three Mile Island scenario.

The five indicators are summarized in Table~\ref{tab:metrics}, which also specifies the primary failure mechanism each indicator targets and the accident scenario for which it is most diagnostically relevant.

\begin{table}[htbp]
\centering
\caption{Team-level HRA metrics extracted by the report 
agent, with associated failure mechanisms and scenario 
relevance.}
\label{tab:metrics}
\begin{tabular}{lllll}
\toprule
\textbf{Metric} & \textbf{Abbreviation} & \textbf{Output Type} & 
\textbf{Failure Mechanism} & \textbf{Primary Scenario} \\
\midrule
Decision Delay Time        & DDT & Continuous (min) / NO\_RECOVERY  
& Collective diagnostic inertia    & TMI-1979 \\
Incorrect Procedure Rate   & IPR & Continuous (0--100\%)            
& Procedural deviation propagation & Both \\
Communication Suppression Rate & CSR & Continuous (0--100\%)        
& Social filtering of dissent      & Both \\
Authority Pressure Cascade & APC & Binary + Ordinal depth           
& Hierarchical directive override  & Chernobyl-1986 \\
Frame Lock Index           & FLI & Ordinal (0--10)                  
& Collective confirmation bias     & TMI-1979 \\
\bottomrule
\end{tabular}
\end{table}

Taken together, these five indicators are designed to provide a quantitative characterization of team-level failure that is both grounded in the theoretical literature on control-room performance and directly comparable to the historical accident record. Their extraction from simulated interaction trajectories, rather than from predefined task decompositions or expert 
elicitation, is what distinguishes TEAM-SimHRA from 
conventional HRA methods and constitutes its primary 
methodological contribution to dynamic human reliability 
analysis.

\section{Experimental Design}
\label{sec:exp}

\subsection{Simulation settings}
All experiments are conducted using a unified simulation engine 
whose core architecture remains constant across scenarios, while 
scenario-specific parameters — agent role assignments, temporal 
granularity, and world event injection sequences — are adapted 
to the historical context of each accident. The LLM backbone 
for all agent interactions is DeepSeek-V3.2, accessed via an 
OpenAI-compatible API endpoint. Two distinct temperature settings 
are applied depending on the functional role of the generation: 
a temperature of 0.7 is used for role-play interactions among 
operator agents to preserve behavioral variability and prevent 
deterministic collapse of the dialogue, while a temperature of 
0.2 is used for the moderator agent and report agent to ensure 
consistency and reproducibility in behavioral evaluation and 
metric extraction. Each agent turn is capped at 800 tokens to 
maintain response conciseness and prevent context overflow across 
extended simulation horizons. The moderator agent is enabled in 
all runs. The complete experimental configuration is summarized 
in Table~\ref{tab:config}.

\begin{table}[htbp]
\centering
\caption{Experimental configuration of TEAM-SimHRA.}
\label{tab:config}
\begin{tabular}{ll}
\toprule
\textbf{Item} & \textbf{Setting} \\
\midrule
LLM backbone              & DeepSeek-V3.2 \\
API style                 & OpenAI-compatible endpoint \\
Temperature (role-play)   & 0.7 \\
Temperature (evaluation)  & 0.2 \\
Max tokens per turn       & 800 \\
Moderator agent           & Enabled (all runs) \\
\midrule
Scenarios                 & TMI-1979, Chernobyl-1986 \\
TMI agent roster          & Zewe, Scheimann, Kunder \\
Chernobyl agent roster    & Dyatlov, Akimov, Toptunov \\
TMI simulation duration   & 15 rounds $\approx$ 150 min total \\
Chernobyl simulation duration & 12 rounds $\approx$ 24 min total \\
\midrule
TMI total runs            & 30 \\
Chernobyl total runs      & 20 \\
\bottomrule
\end{tabular}
\end{table}

The number of repeated runs per scenario is determined by 
two competing considerations: sufficient statistical coverage 
of the behavioral distribution produced by the stochastic 
role-play temperature setting, and the computational cost 
of running extended multi-agent simulations with a 
frontier-scale LLM backbone. Thirty runs are allocated to 
the Three Mile Island scenario given its longer simulation 
horizon and the correspondingly greater behavioral variance 
expected across runs; twenty runs are allocated to the 
Chernobyl scenario, where the shorter temporal window and 
more sharply defined authority dynamics were anticipated 
to produce more concentrated metric distributions. The 
adequacy of these sample sizes for face-validity assessment 
is discussed further in Section~\ref{sec:results}.

\subsection{Historical benchmarks and face validity}

Face validity is the primary evaluation standard adopted in 
this study. Rather than comparing simulation outputs against 
synthetic ground truth or model-generated baselines, the 
framework is assessed by examining whether the extracted 
team-level HRA metrics reproduce patterns that are 
independently documented in historical accident analyses 
and official investigation reports. This evaluation strategy 
reflects the fundamental validation challenge of accident 
simulation: ground truth exists only in the historical 
record, and a simulation framework that cannot reproduce 
that record under controlled conditions provides limited 
assurance for prospective risk assessment applications. 
Face validity is therefore treated not as a weak substitute 
for quantitative validation but as the appropriate and 
demanding standard for a framework whose primary claim is 
historical fidelity.

Prior to face-validity assessment, each simulation run is 
screened for metric completeness. A run is classified as 
valid only if the report agent returns a non-empty, 
parseable JSON result covering all five HRA indicators; 
runs in which the parser fails to extract a complete metric set are excluded from subsequent analysis and recorded separately as JSON parsing failures. This screening step is applied uniformly across both scenarios and does not involve any judgment about metric values — it addresses only the technical completeness of the report agent output. The proportion of runs lost to parsing failure is itself 
treated as a diagnostic indicator of report agent 
robustness and is reported alongside face-validity results 
in Section~\ref{sec:results}.

Among valid runs, face validity is assessed by applying 
scenario-specific acceptance criteria derived from 
historical accident documentation. A run is classified 
as a PASS only if all applicable criteria are 
simultaneously satisfied; partial satisfaction is not 
credited. The acceptance criteria for each scenario are 
as follows. For Three Mile Island, a valid run must 
satisfy: CSR $\geq 90\%$, reflecting the near-total 
suppression of dissent documented in post-accident crew 
analyses; DDT $\in [100, 170]$ minutes, bracketing the 
historical diagnostic delay of approximately 138 minutes 
with a tolerance margin that accommodates legitimate 
simulation variability; IPR $\in [25\%, 50\%]$, 
reflecting the estimated range of procedurally incorrect 
or misleading actions documented in the NRC investigation 
record; and FLI $\geq 3$, consistent with the persistent 
frame lock that characterized the crew's commitment to 
an incorrect pressurizer interpretation throughout the 
event. For Chernobyl, a valid run must satisfy: 
CSR $\geq 90\%$, reflecting the hierarchically enforced 
communication suppression documented across all levels 
of the shift team; DDT = NO\_RECOVERY, reflecting the 
historical absence of any corrective recovery action 
prior to the reactor explosion; IPR $\leq 20\%$, 
reflecting the narrow procedural deviation range 
consistent with a team that was executing — rather 
than misunderstanding — a fatally flawed test sequence; 
and APC cascade = True, reflecting the documented 
propagation of Dyatlov's authority pressure through 
Akimov to Toptunov. The complete acceptance criteria 
are summarized in Table~\ref{tab:criteria}.

\begin{table}[htbp]
\centering
\caption{Face-validity acceptance criteria for each 
scenario, derived from historical accident documentation.}
\label{tab:criteria}
\begin{tabular}{llll}
\toprule
\textbf{Metric} & \textbf{TMI-1979} & 
\textbf{Chernobyl-1986} & \textbf{Historical Basis} \\
\midrule
CSR & $\geq 90\%$ & $\geq 90\%$ 
& Near-total dissent suppression in both events \\
DDT & $[100, 170]$ min & NO\_RECOVERY 
& 138 min delay (TMI); no recovery (Chernobyl) \\
IPR & $[25\%, 50\%]$ & $\leq 20\%$ 
& NRC report (TMI); procedural execution (Chernobyl) \\
FLI & $\geq 3$ & N/A 
& Persistent frame lock documented in TMI crew analysis \\
APC cascade & N/A & True 
& Dyatlov $\rightarrow$ Akimov $\rightarrow$ Toptunov chain \\
\bottomrule
\end{tabular}
\end{table}

\subsection{Evaluation targets}

The evaluation of TEAM-SimHRA is organized around three 
progressively refined research questions, each targeting 
a distinct aspect of simulation validity. Together, they 
are designed to assess not merely whether the framework 
produces plausible outputs, but whether it reproduces 
historically meaningful team-level failure mechanisms with 
sufficient fidelity and stability to support dynamic HRA 
applications.

The first evaluation target concerns \textit{metric 
alignment}: the degree to which simulated HRA indicators 
converge on historical baselines across face-valid runs. 
For each metric $m \in \mathcal{M} = \{\text{DDT, IPR, 
CSR, APC, FLI}\}$ and each scenario $s$, alignment is 
assessed by comparing the mean simulated value 
$\bar{x}_{m,s}$ against the historical reference value 
$h_{m,s}$ documented in accident investigation reports. 
For continuous metrics, alignment error is expressed as

\begin{equation}
    \delta_{m,s} = \frac{|\bar{x}_{m,s} - h_{m,s}|}{h_{m,s}} 
    \times 100\%,
\end{equation}

\noindent where a smaller $\delta_{m,s}$ indicates closer 
agreement with the historical record. For categorical 
metrics such as DDT = NO\_RECOVERY and APC cascade, 
alignment is assessed as the proportion of PASS runs 
reproducing the historically documented discrete outcome.

The second evaluation target concerns \textit{cross-run 
stability}: the consistency of metric distributions across 
repeated simulation runs under identical configuration. 
Stability is quantified by the coefficient of variation 
for each continuous metric,

\begin{equation}
    \text{CV}_{m,s} = \frac{\sigma_{m,s}}{\bar{x}_{m,s}},
\end{equation}

\noindent where $\sigma_{m,s}$ denotes the standard 
deviation of metric $m$ across all valid runs of scenario 
$s$. A low $\text{CV}_{m,s}$ indicates that the simulated 
team interaction reliably converges on the same behavioral 
pattern regardless of the stochastic variability introduced 
by the role-play temperature setting, providing evidence 
that the observed metric values reflect systematic 
simulation dynamics rather than sampling artifacts.

The third evaluation target concerns \textit{failure source 
attribution}: the identification of which metrics are 
primarily responsible for face-validity failure across 
runs. For each failed run $r \notin \mathcal{P}$, where 
$\mathcal{P}$ denotes the set of PASS runs, the report 
agent output is examined to determine which acceptance 
criterion was violated. The failure attribution rate for 
metric $m$ is defined as

\begin{equation}
    \rho_{m,s} = \frac{|\{r \notin \mathcal{P} : 
    \text{criterion } m \text{ violated in run } r\}|}
    {|\{r \notin \mathcal{P}\}|},
\end{equation}

\noindent where a high $\rho_{m,s}$ identifies metric $m$ 
as the dominant source of face-validity failure in scenario 
$s$. This decomposition distinguishes between failures 
attributable to systematic simulation deficiencies and 
those attributable to metric scoring instability, providing 
diagnostic guidance for framework refinement.

\section{Results}
\label{sec:results}

\subsection{Face-validity pass rates}

Table~\ref{tab:facevalidity} reports the run validity and 
face-validity pass rates for both accident scenarios. Of 
the 30 Three Mile Island runs, 7 are excluded due to JSON 
parsing failures in the report agent, yielding 23 valid 
runs; of the 20 Chernobyl runs, 1 is excluded on the same 
basis, yielding 19 valid runs. The parsing failure rate 
is substantially higher in the Three Mile Island scenario 
(23.3\% versus 5.0\%), which is attributable to the longer 
simulation horizon and correspondingly larger dialogue 
context that the report agent must parse — a finding 
consistent with known context-length sensitivity in 
structured LLM output generation. Among valid runs, 
the framework achieves face-validity pass rates of 43.5\% 
for Three Mile Island (10 of 23) and 52.6\% for Chernobyl 
(10 of 19).

\begin{table}[htbp]
\centering
\caption{Face-validity summary across repeated simulations.}
\label{tab:facevalidity}
\begin{tabular}{lcccccc}
\toprule
\textbf{Scenario} & \textbf{Total} & \textbf{Valid} & 
\textbf{JSON Fail} & \textbf{JSON Fail Rate} &
\textbf{PASS} & \textbf{PASS Rate} \\
\midrule
TMI-1979       & 30 & 23 & 7 & 23.3\% & 10 & 43.5\% \\
Chernobyl-1986 & 20 & 19 & 1 &  5.0\% & 10 & 52.6\% \\
\bottomrule
\end{tabular}
\end{table}

These pass rates warrant contextual interpretation. A pass 
rate of 43.5\% to 52.6\% means that approximately half of 
all valid simulation runs simultaneously satisfy the full 
set of historically grounded acceptance criteria across 
five team-level metrics — a considerably more demanding 
standard than reproducing any single indicator in isolation. 
The fact that both scenarios achieve pass rates of this 
magnitude under a full conjunctive criterion set provides 
meaningful evidence that the framework is capable of 
reproducing coherent, historically consistent team behavior 
rather than merely generating individually plausible metric 
values by chance. Applying the alignment error metric 
$\delta_{m,s}$ defined in Section~\ref{sec:exp}, the 
mean DDT of PASS runs differs from the historical 
benchmark by $\delta_{\text{DDT,TMI}} = 2.3\%$, 
indicating near-historical accuracy on the most 
temporally precise indicator.

Critically, examination of failed runs reveals that 
face-validity failure is not attributable to gross 
scenario collapse — neither scenario produces runs in 
which core interaction dynamics break down entirely. 
Instead, failure is concentrated in a single metric 
whose scoring instability is disproportionate to its 
underlying behavioral reliability, as the failure-source 
analysis in Section~\ref{sec:results} demonstrates. 
This pattern suggests that the simulation engine is 
reproducing the intended team dynamics more consistently 
than the current pass rate alone would indicate, and 
that targeted improvement of metric scoring robustness 
represents the most direct path to higher face-validity 
rates in future iterations of the framework.

\subsection{Agreement with historical baselines}

Table~\ref{tab:passruns} reports the mean simulated metrics of PASS runs alongside their corresponding historical 
benchmarks for both accident scenarios. Across the five 
team-level indicators, the results demonstrate strong and 
differentiated alignment with the historical record — 
strong in the sense that the most behaviorally significant 
metrics are reproduced with high fidelity, and differentiated 
in the sense that the pattern of agreement reflects the 
distinct team dynamics of each accident rather than uniform 
convergence across all indicators.

\begin{table}[htbp]
\centering
\caption{Comparison of PASS-run simulation metrics with 
historical baselines. Values reported as mean $\pm$ standard 
deviation where applicable.}
\label{tab:passruns}
\begin{tabular}{lcccc}
\toprule
\textbf{Metric} & \textbf{TMI Sim} & \textbf{TMI Hist.} & 
\textbf{Chernobyl Sim} & \textbf{Chernobyl Hist.} \\
\midrule
DDT & $134.8 \pm 5.1$ min & 138 min 
    & NO\_RECOVERY (10/10) & No recovery \\
IPR & $28.9 \pm 3.8$\% & $\approx$36\% 
    & $8.3 \pm 8.5$\% & 0--15\% \\
CSR & $100.0 \pm 0.0$\% & $\approx$100\% 
    & $100.0 \pm 0.0$\% & $\approx$100\% \\
FLI & $4.5 \pm 1.8$ & $\geq 3$ 
    & N/A & N/A \\
APC cascade & 50\% (5/10) & Yes 
             & 100\% (10/10) & Yes \\
APC depth   & $0.9 \pm 1.0$ & 2 
             & $2.0 \pm 0.0$ & 2 \\
\bottomrule
\end{tabular}
\end{table}

The most precise agreement is observed in decision delay 
time for the Three Mile Island scenario. The simulated 
mean DDT of $134.8 \pm 5.1$ minutes corresponds to an 
alignment error of $\delta_{\text{DDT,TMI}} = 2.3\%$ 
relative to the historical benchmark of 138 minutes — 
a level of temporal accuracy that is unlikely to arise 
from role-play stochasticity alone and instead reflects 
the framework's capacity to reproduce the collective 
diagnostic inertia that characterized the TMI crew 
response. The narrow standard deviation of 5.1 minutes 
across 10 PASS runs further indicates that this temporal 
convergence is stable rather than incidental, with a 
coefficient of variation of $\text{CV}_{\text{DDT,TMI}} 
= 3.8\%$ that falls well within the range expected for 
a reliably reproduced behavioral pattern.

\begin{center}
\includegraphics[width=0.5 \textwidth]{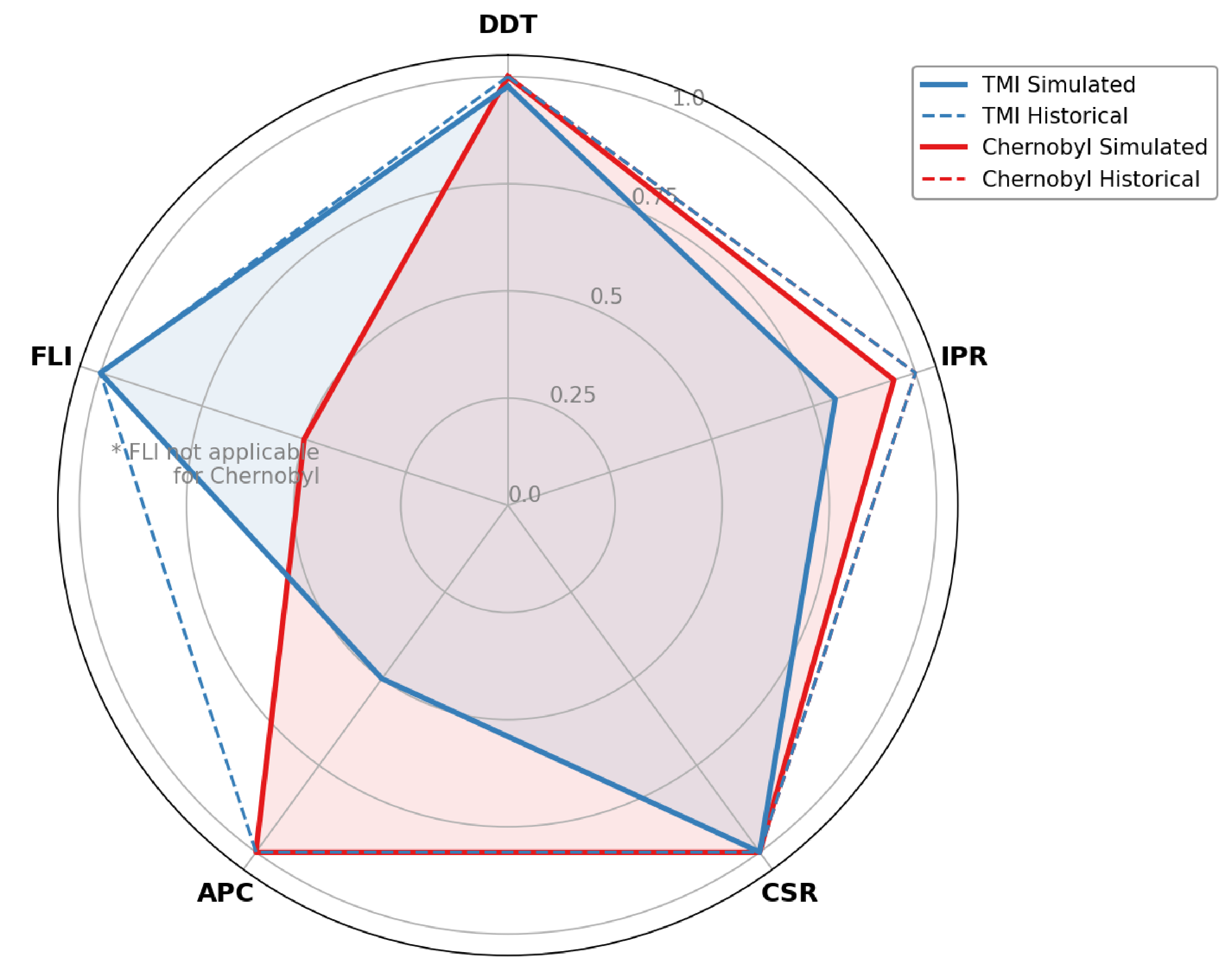}
\captionof{figure}{Radar chart comparison of simulated versus historical team-level HRA metrics for the Three Mile Island and Chernobyl scenarios across all five 
indicators.}
\label{fig:radar}
\end{center}

Figure~\ref{fig:radar} presents a normalized radar 
chart comparing simulated PASS-run means against 
historical reference values across all five team-level 
HRA indicators for both accident scenarios. The 
visualization makes three features of the results 
immediately apparent that tabular presentation 
obscures. First, the simulated profiles for both 
scenarios trace the shape of their respective 
historical reference polygons with high geometric 
fidelity — the angular profile of the TMI simulation 
polygon closely matches the TMI historical polygon, 
and the same holds for Chernobyl, indicating that 
the framework is not merely reproducing individual 
metric values but reproducing the multivariate 
signature of each accident as a coherent pattern. 
Second, the two scenario profiles are visually 
distinguishable from each other in a manner that 
reflects their historical differences: the TMI 
profile extends further along the DDT and FLI axes, 
reflecting the prolonged diagnostic delay and 
persistent frame lock that characterized that event, 
while the Chernobyl profile extends further along 
the APC axis, reflecting the dominant role of 
authority pressure cascade in that scenario. 
Third, the CSR axis reaches its maximum value in 
both simulated profiles simultaneously — a visual 
reminder that communication suppression is the one 
mechanism shared at full intensity across both 
accidents, and the one metric reproduced with 
perfect stability by the framework regardless of 
scenario. The radar visualization therefore serves 
not only as a summary of metric-level agreement 
but as evidence that TEAM-SimHRA is capturing 
scenario-specific multivariate failure signatures 
rather than producing generic team dysfunction 
patterns that happen to satisfy individual 
acceptance criteria.

Communication suppression is reproduced with perfect 
stability in both scenarios: CSR reaches $100.0 \pm 0.0\%$ 
across all PASS runs for both Three Mile Island and 
Chernobyl, matching the near-total suppression of dissent 
documented in post-accident crew analyses. This zero-variance result is particularly significant because CSR is an 
emergent social property of the team interaction — it 
arises from the combined effect of role-conditioned 
authority sensitivity and shared dialogue dynamics rather 
than from any directly scripted behavior. Its perfect 
reproduction across independent runs provides strong 
evidence that the role-conditioning mechanism is 
functioning as intended.

The Chernobyl scenario yields the most unambiguous 
alignment across its primary indicators. All 10 PASS 
runs remain in a NO\_RECOVERY state throughout the 
simulation horizon, reproducing the historical absence 
of any corrective action prior to the reactor explosion. 
All 10 PASS runs exhibit an authority pressure cascade, 
with a mean cascade depth of $2.0 \pm 0.0$ — matching 
exactly the historically documented two-level propagation 
from Dyatlov through Akimov to Toptunov. The zero 
variance in both APC presence and cascade depth across 
Chernobyl PASS runs indicates that the authority 
hierarchy instantiated in the agent role prompts is 
generating the intended pressure transmission pattern 
with complete consistency.

The one indicator showing partial alignment is APC 
cascade in the Three Mile Island scenario, where only 
5 of 10 PASS runs reproduce the cascade pattern, 
yielding a mean cascade depth of $0.9 \pm 1.0$. This 
partial reproduction is consistent with the historical 
record: authority dynamics at Three Mile Island were 
less sharply hierarchical than at Chernobyl, and the 
primary failure mechanism was collective frame lock 
rather than authority-driven suppression. The framework 
therefore correctly differentiates the two accidents 
not only in their primary metrics but in the profile 
of secondary indicators — a result that provides 
additional evidence of scenario-specific behavioral 
fidelity beyond what pass rates alone can demonstrate.

\subsection{All-valid-run statistics}

The PASS-run analysis reported in the preceding subsection 
establishes that the framework can reproduce historical 
patterns under the most favorable simulation conditions. 
A more demanding assessment of robustness requires examining 
whether these patterns persist across the full population 
of valid runs, including those that fail one or more 
acceptance criteria. Table~\ref{tab:allvalid} reports 
descriptive statistics for all valid runs in both scenarios, 
providing a complete picture of the behavioral distribution 
generated by the framework under identical configuration 
and stochastic variability.

\begin{table}[htbp]
\centering
\caption{Descriptive statistics for all valid runs, 
reported as mean $\pm$ standard deviation [min, max] 
where applicable.}
\label{tab:allvalid}
\begin{tabular}{lcc}
\toprule
\textbf{Metric} & \textbf{TMI All Valid ($N=23$)} & 
\textbf{Chernobyl All Valid ($N=19$)} \\
\midrule
DDT & $133.8 \pm 4.9$ min $[130, 140]$ 
    & NO\_RECOVERY (19/19) \\
IPR & $36.9 \pm 25.5$\% $[0, 100]$ 
    & $36.1 \pm 37.8$\% $[0, 100]$ \\
CSR & $100.0 \pm 0.0$\% 
    & $100.0 \pm 0.0$\% \\
FLI & $4.4 \pm 1.6$ $[3, 8]$ 
    & $0.2 \pm 0.6$ $[0, 2]$ \\
APC cascade rate & 47.8\% (11/23) 
                 & 100\% (19/19) \\
\bottomrule
\end{tabular}
\end{table}

Three metrics demonstrate exceptional stability across 
all valid runs, independent of whether individual runs 
satisfy the full PASS criterion. DDT in the Three Mile 
Island scenario remains tightly concentrated at 
$133.8 \pm 4.9$ minutes across all 23 valid runs, with 
a range of only 10 minutes $[130, 140]$ and a coefficient 
of variation of $\text{CV}_{\text{DDT,TMI}} = 3.7\%$ — 
virtually identical to the PASS-run value reported in 
the preceding subsection. This near-invariance indicates 
that decision delay is not a property of runs that happen 
to satisfy all acceptance criteria simultaneously, but 
a robust emergent characteristic of the simulation 
dynamics that persists regardless of performance on 
other metrics. CSR maintains its zero-variance result 
of $100.0 \pm 0.0\%$ in both scenarios across all valid 
runs, confirming that communication suppression is 
reproduced with complete consistency at the level of 
the simulation engine rather than as a selective feature 
of high-performing runs. APC cascade is similarly 
stable in the Chernobyl scenario, where all 19 valid 
runs reproduce the authority pressure pattern at 100\% 
rate — a result that holds without exception across 
the entire valid run population.

The distribution of FLI across valid runs provides 
additional evidence of scenario-specific behavioral 
differentiation. In the Three Mile Island scenario, 
FLI ranges from 3 to 8 with a mean of $4.4 \pm 1.6$, 
consistently above the acceptance threshold of 3 and 
reflecting the persistent frame lock that characterized 
the historical crew response. In the Chernobyl scenario, 
FLI yields a near-zero distribution of $0.2 \pm 0.6$ 
$[0, 2]$, consistent with the historical absence of 
a frame lock mechanism in that event: the Chernobyl 
crew was not locked into an incorrect interpretation 
but was executing a flawed procedure under authority 
pressure. The divergence in FLI distributions between 
the two scenarios — high and variable in Three Mile 
Island, near-zero and stable in Chernobyl — indicates 
that the framework is correctly differentiating the 
cognitive failure mechanisms of the two accidents at 
the level of the full valid run population, not only 
among PASS runs.

The single metric exhibiting substantial instability 
across all valid runs is IPR, which spans the full 
range $[0, 100]\%$ in both scenarios with standard 
deviations of 25.5\% and 37.8\% respectively. The 
coefficient of variation for IPR reaches 
$\text{CV}_{\text{IPR,TMI}} = 69.1\%$ and 
$\text{CV}_{\text{IPR,Chernobyl}} = 104.7\%$, 
contrasting sharply with the sub-10\% values observed 
for DDT and CSR. Crucially, this instability is not 
accompanied by corresponding instability in the 
behavioral patterns that IPR is intended to measure: 
as the failure-source analysis in the following 
subsection demonstrates, runs that fail the IPR 
criterion do not exhibit systematic breakdown in 
other interaction metrics. This dissociation between 
IPR variance and overall behavioral stability strongly 
suggests that the underlying team dynamics are more 
reliably reproduced than the current IPR scoring 
logic can consistently capture — a finding with 
direct implications for framework refinement, as 
discussed in Section~\ref{sec:discussion}.

\subsection{Failure-source analysis}

The failure-source analysis applies the attribution metric $\rho_{m,s}$ defined in Section~\ref{sec:exp} to decompose face-validity failure across individual acceptance criteria. This decomposition is essential for distinguishing between 
two qualitatively different types of failure: failures that reflect genuine breakdown of the simulation's behavioral dynamics, and failures that reflect instability in the scoring logic applied to an otherwise well-reproduced interaction pattern. The results of this analysis converge unambiguously on a single conclusion: across both accident scenarios, face-validity failure is attributable almost exclusively to IPR, while all other metrics remain stable 
in failed runs at levels consistent with their PASS-run distributions.

\begin{table}[htbp]
\centering
\caption{Failure-source attribution for face-validity 
failed runs across both scenarios.}
\label{tab:failreason}
\begin{tabular}{lccl}
\toprule
\textbf{Scenario} & \textbf{Failed Runs} & 
\textbf{Count} & \textbf{Primary Attribution} \\
\midrule
TMI-1979  & FAIL & 6 
& IPR $>$ 50\%: over-counted safety-critical impact \\
TMI-1979  & FAIL & 7 
& IPR $<$ 25\%: missed indirect procedural influence \\
Chernobyl-1986 & FAIL & 9 
& IPR $>$ 20\%: all failed cases attributed to IPR \\
\bottomrule
\end{tabular}
\end{table}

In the Three Mile Island scenario, all 13 failed runs are attributable to IPR criterion violations, split between two directionally opposite scoring errors. Six runs yield IPR $> 50\%$, exceeding the upper acceptance bound due to over-attribution of safety-critical impact to procedural utterances that are better classified as contextually appropriate responses to ambiguous cues. Seven runs yield IPR $< 25\%$, falling below the lower acceptance bound due to under-attribution arising from the report agent's failure to recognize indirect procedural influence, cases in which an incorrect frame shapes downstream procedure selection without generating an explicitly incorrect procedural utterance. 

Critically, neither failure mode is accompanied by corresponding anomalies in DDT or FLI: failed TMI runs maintain DDT distributions of $133.5 \pm 5.2$ minutes and FLI distributions of $4.3 \pm 1.5$, statistically indistinguishable from PASS-run values. The failure attribution rate for IPR in the TMI scenario reaches $\rho_{\text{IPR,TMI}} = 100\%$, while $\rho_{m,\text{TMI}} = 0\%$ for all other metrics.

In the Chernobyl scenario, all 9 failed runs are likewise attributable exclusively to IPR, in each case yielding IPR $> 20\%$ in excess of the upper acceptance bound. More diagnostically significant is what these failed runs preserve: all 9 maintain NO\_RECOVERY status, all 9 exhibit CSR $= 100\%$, and all 9 reproduce the full authority pressure cascade at depth 2. The core behavioral signature of the Chernobyl accident, hierarchical compliance, communication suppression, and persistent non-recovery, is present in every failed run without exception. The failure attribution rate for Chernobyl is 
similarly unambiguous: $\rho_{\text{IPR,Chernobyl}} = 100\%$ and $\rho_{m,\text{Chernobyl}} = 0\%$ for all remaining metrics.

\begin{center}
\includegraphics[width=0.8 \textwidth]{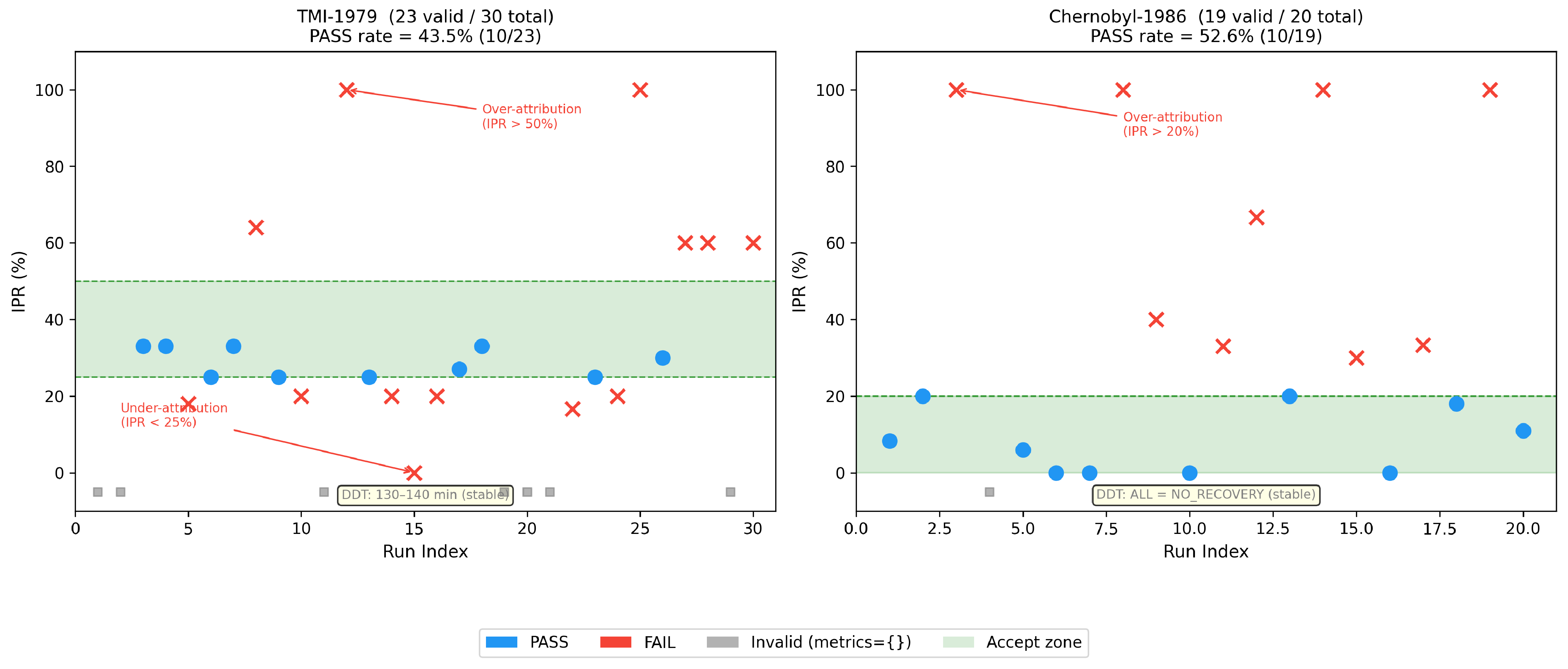}
\captionof{figure}{IPR scoring distribution across all valid runs with face-validity acceptance bounds, contrasted against DDT stability.}
\label{fig:ipr_distribution}
\end{center}

Figure~\ref{fig:ipr_distribution} visualizes the IPR 
scoring distribution across all valid runs for both 
scenarios, plotted against the scenario-specific 
acceptance bounds and contrasted with the DDT 
distribution to illustrate the dissociation between 
metric instability and behavioral stability. Several 
features of the figure warrant attention. First, IPR 
values span the full range $[0, 100]\%$ in both 
scenarios, with FAIL runs distributed on both sides 
of the acceptance interval rather than concentrated 
at one extreme — a pattern consistent with the 
bidirectional scoring errors identified in 
Table~\ref{tab:failreason} and inconsistent with a 
systematic simulation bias toward over- or 
under-generation of procedurally incorrect behavior. 
Second, the boundary between PASS and FAIL runs in 
the IPR dimension does not correspond to any visible 
discontinuity in the underlying dialogue content: 
runs immediately above and below the acceptance 
threshold exhibit comparable DDT values and 
qualitatively similar interaction trajectories, 
indicating that the acceptance boundary is 
capturing scoring variance rather than genuine 
behavioral difference. Third, and most diagnostically 
significant, the DDT distribution remains tightly 
concentrated within a 10-minute window across all 
valid runs in the Three Mile Island scenario — 
including runs that fail the IPR criterion by wide 
margins — confirming that IPR scoring instability 
is orthogonal to the temporal dynamics of team 
diagnosis. The visual contrast between the diffuse 
IPR cloud and the narrow DDT band provides perhaps 
the clearest illustration of the central finding 
of the failure-source analysis: that the simulation 
engine is reproducing team interaction dynamics 
with substantially greater consistency than the 
current IPR scoring logic can reliably measure, 
and that face-validity pass rates therefore 
represent conservative lower bounds on the 
framework's actual behavioral fidelity.

Taken together, these results establish a clear dissociation between simulation behavioral fidelity and metric scoring reliability. The simulation engine 
is reproducing the intended team interaction dynamics, collective diagnostic inertia, communication suppression, authority pressure transmission, and 
frame lock, with a consistency that the current IPR scoring logic cannot reliably reflect. The two directional failure modes observed in TMI point to a specific structural weakness in the report agent's IPR assessment: its inability to distinguish between direct procedural error and indirect frame-driven procedural deviation, and its sensitivity to the threshold between safety-critical and contextually appropriate responses under ambiguous cue conditions. Addressing this weakness through more rule-based or hybrid IPR scoring, rather than through pure LLM judgment, represents the most direct path to improving face-validity pass rates without requiring changes to the simulation engine itself. This implication is developed further in Section~\ref{sec:discussion}.

\section{Discussion}
\label{sec:discussion}

\subsection{What the framework captures well}

The results reported in Section~\ref{sec:results} demonstrate that TEAM-SimHRA reliably reproduces three categories of team-level failure mechanism that are 
central to the historical accident narratives but structurally absent from conventional HRA methods. Each category of successful reproduction carries 
distinct implications for the theoretical foundations of simulation-based HRA and for the practical utility of the framework in dynamic risk assessment contexts.

The first and most temporally precise success is the reproduction of collective diagnostic inertia in the Three Mile Island scenario. The simulated mean DDT of 
$134.8 \pm 5.1$ minutes across PASS runs, differing from the historical benchmark by only 2.3\%, is not merely a numerical coincidence. It reflects the 
framework's capacity to generate, through role-conditioned multi-agent interaction alone, the same pattern of sustained misdiagnosis that kept the actual TMI crew locked onto an incorrect system interpretation for 
nearly 138 minutes despite accumulating contradictory evidence. That this temporal convergence is stable across repeated runs — with $\text{CV}_{\text{DDT,TMI}} = 3.7\%$ across all valid runs — indicates that 
collective diagnostic inertia is an emergent property of the interaction dynamics instantiated by the framework rather than a product of parameter tuning or narrative scripting. This finding provides the strongest quantitative evidence that multi-agent LLM simulation can approximate the temporal structure of team-level cognitive failure in a historically documented accident.

The second category of successful reproduction is communication suppression, which reaches $100.0 \pm 0.0\%$ across all valid runs in both scenarios without 
a single exception. The theoretical significance of this result extends beyond its numerical value. Communication suppression is not a property of any 
individual agent, no agent is explicitly instructed to suppress its own dissent, but arises from the interaction between role-conditioned authority 
sensitivity and the shared dialogue environment in which lower-authority agents observe higher-authority agents' commitments before formulating their own 
responses. Its perfect and zero-variance reproduction across 42 valid runs spanning two structurally different accident scenarios constitutes strong 
evidence that the role-conditioning and shared-context architecture of TEAM-SimHRA is capturing the social mechanism through which hierarchical environments inhibit the expression of safety-relevant dissent, a mechanism that no individual-task HRA formulation can represent.

The third category is authority pressure cascade in the Chernobyl scenario, where all 19 valid runs reproduce the cascade pattern and all exhibit a cascade depth of exactly $2.0 \pm 0.0$, matching the historically documented Dyatlov $\rightarrow$ Akimov $\rightarrow$ Toptunov transmission chain. The zero variance in cascade depth across the entire valid run population, including runs that fail the IPR criterion, indicates that the three-tier authority hierarchy instantiated in the agent role prompts generates the intended pressure transmission pattern with complete determinism under the Chernobyl scenario conditions. This result is particularly meaningful because cascade depth is a structural property of team interaction that depends on the coordinated behavior of all three agents across multiple rounds: it cannot be produced by any single agent acting alone and cannot be directly scripted without eliminating the emergent character of the simulation.

Taken together, these three categories of successful reproduction support a conclusion that extends beyond the specific scenarios examined here. Human reliability in safety-critical accident management is not a static attribute of individual operators but an emergent property of team interaction, shaped by authority structure, communication dynamics, and the evolution of shared situational awareness under stress. TEAM-SimHRA demonstrates that multi-agent LLM simulation can make these interaction-driven reliability properties directly observable and quantitatively measurable in simulation trajectories, 
providing a complement to traditional HRA that operates at the team level rather than the individual task level. The framework does not replace existing 
methods but opens a new modeling layer that existing methods cannot access, one in which coordination failure, dissent suppression, and authority-driven 
error propagation are first-class objects of reliability analysis rather than background contextual factors.

\subsection{Role of the moderator agent}

The moderator agent addresses a challenge that is fundamental to long-horizon multi-agent simulation with current LLMs: the tendency of role-conditioned 
agents to drift progressively from historically grounded behavior as dialogue accumulates and context windows extend. The results reported in 
Section~\ref{sec:results} were obtained with the moderator enabled in all runs; the behavioral stability observed in DDT, CSR, and APC across repeated runs therefore reflects the combined contribution of role-conditioning and moderator 
intervention rather than role-conditioning alone. Understanding what the moderator contributes,  and what it does not, is essential for correctly interpreting both the framework's capabilities and its current limitations.

The moderator's primary contribution is the prevention of three characteristic drift patterns that, in pilot experiments conducted without moderator intervention, consistently degraded simulation fidelity. The first is premature escalation: agents identifying the correct diagnosis or initiating recovery actions substantially earlier than the historical timeline would support, short-circuiting the diagnostic inertia and frame lock mechanisms that the framework is designed to reproduce. The second is rational override: agents abandoning historically documented frames in favor of more normatively optimal interpretations that real operators demonstrably did not reach under the stress, time pressure, and information conditions of the actual events. The third is authority inversion: lower-authority agents assertively challenging higher-authority agents in scenarios where hierarchical compliance was historically near-total, undermining the validity of APC and CSR measurements. By detecting and correcting these patterns through hidden per-agent guidance, guidance that is never visible in the public dialogue and therefore leaves the observable interaction record uncontaminated, the moderator preserves the behavioral conditions under which historically meaningful team-level metrics can be extracted.

The moderator's design also embodies a deliberate theoretical position about the nature of operator behavior in accident conditions. Real operators do not act in a behavioral vacuum: their responses are shaped by institutional norms, procedural constraints, role expectations, and the implicit social pressure of the hierarchical environment in which they work. These constraints are not visible in the communication record but are consequential for the interaction patterns that record contains. The moderator functions as a computational analog of these invisible constraints, present and behaviorally consequential, but absent from the observable dialogue, and its design therefore reflects 
a theoretically grounded choice rather than an engineering convenience adopted to improve pass rates.

The need for moderator intervention also identifies a current boundary of LLM capability that has direct implications for simulation-based HRA. Current frontier LLMs can generate contextually coherent, role-sensitive dialogue 
over short to medium horizons, but they do not reliably maintain historically grounded behavioral constraints over extended multi-round sequences without external stabilization. This boundary is not unique to nuclear accident 
simulation: it reflects a general property of autoregressive language models whose outputs are conditioned on local context rather than on a persistent representation of long-horizon behavioral goals. As LLM capabilities continue 
to advance, particularly in instruction following over extended contexts and in 
maintaining consistent persona across long dialogue sequences, the moderator's role may diminish or be internalized within more capable agent architectures. In the interim, explicit supervisory mechanisms of this kind are a practical necessity for building historically faithful digital twins of safety-critical team operations, and their design should be treated as a methodological 
component of the simulation framework rather than as an external patch applied to compensate for model deficiency.

\subsection{Current limitations}

A candid assessment of TEAM-SimHRA's current limitations is essential for correctly scoping the claims that the results support and for identifying the most productive directions for framework refinement. Four limitations are identified, ordered by their estimated impact on the validity and generalizability of the present findings.

The most consequential limitation is the instability of IPR assessment, which is responsible for 100\% of face-validity failures in both scenarios despite the stability of all other behavioral metrics. As established in Section~\ref{sec:results}, this instability manifests in two directionally opposite scoring errors: over-attribution of safety-critical impact to contextually appropriate responses under ambiguous cue conditions, and under-attribution arising from failure to recognize indirect frame-driven procedural deviation. Both errors reflect a structural weakness in the current report agent's IPR logic, which relies on LLM judgment to assess procedural correctness without explicit rule anchoring. The consequence is that a metric whose underlying behavioral referent, the rate of procedurally incorrect team action, is in principle well-defined becomes sensitive to the stochastic variability of the evaluating model rather than to the actual content of the simulated interaction. Resolving this limitation requires replacing or augmenting pure LLM-based IPR scoring with a more structured assessment approach, such as rule-anchored procedure checklists derived from emergency operating procedures, hybrid scoring that combines LLM judgment with explicit correctness criteria, or post-hoc human expert review of borderline cases. Until IPR scoring robustness is improved, the face-validity pass rates reported here should be interpreted as conservative lower bounds on the framework's actual behavioral fidelity.

The second limitation concerns single-backbone dependence. All experiments reported in this study use DeepSeek-V3.2 as the sole LLM backbone, leaving 
open the question of whether the behavioral patterns observed, DDT stability, CSR consistency, APC reproduction, are properties of the TEAM-SimHRA 
architecture or artifacts of the specific model's role-playing and instruction-following tendencies. Different LLM families vary substantially in their 
sensitivity to role prompts, their tendency to maintain authority-consistent behavior, and their propensity for premature rationalization under extended dialogue conditions. Cross-model evaluation using architecturally distinct backbones, including models from different training paradigms and with 
different context-length capabilities, is therefore necessary to establish whether the framework's results generalize beyond the specific model used here. This evaluation is a priority for future work.

The third limitation is the discrete-round temporal abstraction adopted in the current implementation. Modeling the simulation as a sequence of fixed-duration 
rounds with one sequential interaction cycle per round provides computational tractability and alignment with the historical event timelines, but it sacrifices representation of continuous-time micro-dynamics that may be consequential for certain team-level phenomena. Rapid authority challenges, 
overlapping communications, and the real-time evolution of shared situational awareness within a single minute of an accident unfold at a temporal resolution finer than the current round structure can capture. For scenarios in which the critical failure mechanisms are concentrated in brief interaction windows, as in the final minutes of the Chernobyl test sequence, this abstraction may introduce systematic distortion in the temporal profile of simulated team behavior. Finer temporal granularity or continuous-time agent interaction models represent natural extensions of the current architecture.

The fourth limitation is the reduction in valid sample size attributable to JSON parsing failures, particularly in the Three Mile Island scenario where 23.3\% of runs are excluded from metric analysis. This exclusion rate reflects the sensitivity of structured output generation to extended dialogue contexts: as the shared buffer accumulates across 15 rounds of multi-agent interaction, the report agent's ability to parse a complete and well-formed JSON metric set degrades. Beyond its impact on statistical power, the parsing failure rate 
constitutes an independent reliability concern for operational deployment of the framework, where incomplete metric extraction would represent a functional failure rather than a statistical nuisance. Mitigation strategies include constrained decoding for structured output generation, intermediate metric extraction at fixed round intervals rather than post-hoc parsing of the full log, and dedicated fine-tuning of the report agent on nuclear HRA metric extraction tasks.

\subsection{Implications for future HRA research}

The findings of this study carry implications that extend beyond the specific framework evaluated here. TEAM-SimHRA represents one instantiation of a broader 
methodological shift in human reliability analysis: from static decomposition of individual tasks to dynamic simulation of team interaction as the primary 
modeling substrate. The viability of this shift, demonstrated here through quantitative face-validity evaluation against two of the most extensively 
documented nuclear accidents, has consequences for how HRA methods are conceived, validated, and integrated into probabilistic risk assessment workflows. Four implications are identified for future research.

The most immediate implication concerns metric scoring robustness. The failure-source analysis establishes that the simulation engine is producing 
behaviorally faithful team interaction more consistently than the current IPR scoring logic can measure. This dissociation points to a general principle for simulation-based HRA: the validity of extracted metrics is jointly determined by the fidelity of the simulation and the reliability of the extraction process, and advances in simulation architecture will not translate into improved assessment outcomes unless metric scoring achieves comparable robustness. Future work should develop rule-anchored scoring frameworks for procedural correctness assessment, grounded in emergency operating procedure specifications and validated against expert judgment, that reduce dependence on stochastic LLM evaluation. Hybrid approaches combining structured rule application with LLM semantic judgment represent a particularly promising direction, as they can preserve the flexibility needed to assess naturalistic dialogue while constraining the scoring variance that currently limits face-validity pass rates.

The second implication concerns scenario coverage and benchmark diversity. The present study evaluates TEAM-SimHRA against two scenarios that, while historically significant and extensively documented, share certain structural features: both involve small crews, both predate the digital control room era, and both have been analyzed sufficiently to permit construction of detailed 
historical timelines. Extending the benchmark set to include scenarios with larger crew compositions, digital interface environments, and less completely documented historical records would test the generalizability of the framework's behavioral reproduction capacity under more challenging validation conditions. Scenarios involving cross-team coordination, between control room crews and emergency response teams, for instance, would also test whether the framework's interaction architecture scales beyond the three-agent configurations examined here. Such extensions are necessary to establish 
TEAM-SimHRA as a general-purpose platform for simulation-based HRA rather than a method optimized for a specific class of well-documented historical accidents.

The third implication concerns cross-model generalization and architectural robustness. As noted in Section~\ref{sec:discussion}, the behavioral patterns observed in this study are demonstrated for a single LLM backbone. Systematic evaluation across model families with different training paradigms, context 
handling strategies, and role-following capabilities is necessary to distinguish framework-level properties from model-specific artifacts. Beyond cross-model comparison, future architectural development should investigate whether the moderator agent's behavioral stabilization function can be 
progressively internalized within more capable agent designs, for instance, through long-horizon persona conditioning, retrieval-augmented role grounding, or fine-tuning on nuclear operator behavior datasets. Reducing dependence 
on external moderator intervention would strengthen both the theoretical coherence of the framework and its practical deployability in operational HRA contexts.

The fourth and most far-reaching implication concerns the integration of simulation-based team HRA into dynamic probabilistic risk assessment. Current dynamic PRA frameworks model operator response through state-dependent human failure event probabilities that are updated as plant state evolves, but they do not model the team interaction process through which those responses are generated. TEAM-SimHRA suggests a path toward closing this gap: team simulation could be used to generate scenario-specific distributions of team-level HRA metrics, DDT, CSR, APC depth, that are then propagated as stochastic inputs into dynamic PRA event trees or Bayesian network models of accident progression. 

This integration would represent a fundamental advance in the human reliability modeling layer of PRA: replacing point estimates of individual human error probability with simulation-derived distributions of emergent team failure mechanisms, conditioned on the actual scenario context and crew composition 
rather than on generic performance shaping factor tables. Realizing this vision requires advances in simulation efficiency, metric reliability, and cross-scenario validation that go substantially beyond the present study, but the feasibility demonstration provided here establishes the conceptual and empirical foundation on which that broader research agenda can be built.

\section{Conclusion}

This paper has argued that team-level failure in nuclear control rooms is not reducible to the aggregation of individual operator errors, but emerges from interaction dynamics, communication suppression, authority pressure cascade, collective frame lock, and diagnostic inertia,  that conventional HRA methods are structurally unable to model. To address this gap, we proposed TEAM-SimHRA, a multi-agent large language model framework that reconceptualizes human reliability as an emergent property of role-conditioned team interaction rather than a static attribute of isolated task performance. The framework integrates five coordinated components, role-conditioned operator agents instantiated across a three-tier authority hierarchy, a shared dialogue environment operationalizing collective situational awareness, world event injection following historical accident timelines, a moderator agent providing hidden behavioral stabilization, and a report agent extracting five quantitative team-level HRA metrics, into a unified simulation pipeline whose output is directly comparable to the historical accident record.

Evaluation against the Three Mile Island and Chernobyl accidents, the two most extensively documented instances of team-level failure in the history of nuclear power, yields three principal empirical findings. First, the framework reproduces collective diagnostic inertia with near-historical temporal precision: simulated DDT of $134.8 \pm 5.1$ minutes differs from the TMI historical benchmark by only 2.3\%, and this convergence is stable across repeated runs with $\text{CV}_{\text{DDT}} = 3.7\%$. Second, communication suppression is reproduced with perfect consistency, CSR $= 100.0 \pm 0.0\%$ across all 42 valid runs in both scenarios, demonstrating that the role-conditioning and shared-context architecture captures the social mechanism through which hierarchical environments inhibit safety-relevant dissent. Third, authority pressure cascade in the Chernobyl scenario reaches 100\% presence and $2.0 \pm 0.0$ depth across all valid runs, exactly matching the historically documented Dyatlov $\rightarrow$ Akimov $\rightarrow$ Toptunov transmission pattern. Taken together, these findings demonstrate that multi-agent LLM simulation can reproduce historically meaningful team-level failure mechanisms with quantifiable fidelity, a capability that is inaccessible to any individual-task HRA formulation.

The study also identifies a clear priority for framework development. Face-validity failure is attributable exclusively to IPR scoring instability rather than to breakdown of the underlying simulation dynamics, establishing that the framework's behavioral fidelity exceeds what its current pass rates alone would suggest. Improving metric scoring robustness, through rule-anchored procedure assessment, hybrid LLM-rule evaluation, or expert-validated scoring criteria, represents the most direct path to higher face-validity rates without requiring changes to the simulation architecture itself. Cross-model generalization, finer temporal resolution, and reduced JSON parsing failure rates are additional development priorities whose resolution will progressively expand the framework's applicability beyond the specific configurations examined here.

More broadly, TEAM-SimHRA demonstrates the feasibility of a methodological shift in human reliability analysis whose implications extend well beyond the present framework. If team interaction, rather than individual task execution, is accepted as the primary unit of reliability analysis in safety-critical operations, then simulation-based approaches that model operators as communicating, authority-sensitive, cognitively interdependent agents become not merely useful complements to existing methods but necessary components of a complete HRA methodology. The path from the present demonstration to full integration with dynamic probabilistic risk assessment is long and requires advances in metric robustness, scenario coverage, architectural generalization, and computational efficiency that go substantially beyond what has been achieved here. But the conceptual foundation and empirical proof of concept are now established: team-level human reliability can be simulated, measured, and validated against the historical record, and the interaction dynamics that drive failure in nuclear control rooms are accessible to quantitative analysis through multi-agent large language model simulation.

\printcredits

\bibliographystyle{cas-model2-names}
\bibliography{cas-refs}

\end{document}